\documentclass[a4,11pt]{article}
\usepackage{latexsym}
\usepackage{amssymb}
\usepackage{amsmath}
\usepackage{amsfonts}
\usepackage{amsthm}
\usepackage{url}
\usepackage{epsf}
\usepackage{graphicx,multicol}
\usepackage{epic,eepic}

\begin{document}
\newcommand{\2}{\hspace{0.2 cm}}
\newcommand{\1}{\hspace{0.1 cm}}
\newcommand{\dom}{\mbox{$\rightarrow$}}
\newcommand{\pf}{{\bf Proof: }}
\newtheorem{theorem}{Theorem}[section]
\newtheorem{proposition}[theorem]{Proposition}
\newtheorem{lemma}[theorem]{Lemma}
\newtheorem{problem}[theorem]{Problem}
\newtheorem{definition}[theorem]{Definition}
\newtheorem{corollary}[theorem]{Corollary}
\newtheorem{conjecture}[theorem]{Conjecture}
\newcommand{\<}[1]{\left\langle{#1}\right\rangle}
\newcommand{\beq}{\begin{equation}}
\newcommand{\eeq}{\end{equation}}
\newcommand{\induce}[2]{#1\langle{}#2\rangle}
\newcommand{\dist}{{\rm dist}}
\newcommand{\diam}{{\rm diam}}
\newcommand{\domn}{{\rm domn}}
\newcommand{\vecc}[1]{\stackrel{\leftrightarrow}{#1}}
\newcommand{\ct}{\!\cdot\!}
\newcommand{\bm}[1]{\mbox{\boldmath #1}}
\newcommand{\pc}{{\rm pc}}
\newcommand{\ra}{\rangle}
\newcommand{\la}{\langle}

\title{Properly Coloured Cycles and Paths: Results and Open Problems}

\author{
 Gregory Gutin\thanks{Department of Computer Science,
 Royal Holloway, University of London,
Egham, Surrey TW20 0EX, UK, {\tt gutin@cs.rhul.ac.uk}} \and Eun Jung
Kim\thanks{Department of Computer Science,
 Royal Holloway, University of London,
Egham, Surrey TW20 0EX, UK, {\tt eunjung@cs.rhul.ac.uk}} }

\date{}
\maketitle
\bibliographystyle{plain}
\begin{abstract}
In this paper, we consider a number of results and seven conjectures
on properly edge-coloured (PC) paths and cycles in edge-coloured
multigraphs. We overview some known results and prove new ones. In
particular, we consider a family of transformations of an
edge-coloured multigraph $G$ into an ordinary graph  that allow us to
check the existence PC cycles and PC $(s,t)$-paths in $G$ and, if
they exist, to find shortest ones among them. We raise a problem of
finding the optimal transformation and consider a possible solution
to the problem.
\end{abstract}

\section{Introduction}

The class of edge-coloured multigraphs generalize directed graphs. There
are several other generalizations of directed graphs such as arc-coloured
digraphs, hypertournaments and star hypergraphs, but the class of
edge-coloured multigraphs has been given the main attention in graph
theory literature because many concepts and results on directed
graphs can be extended to edge-coloured multigraphs and there are several
important applications of edge-coloured multigraphs. For a more extensive
treatment of this topic, see \cite{bang2000,bang2009}.

In this paper we overview some known results on properly coloured
(PC) cycles and paths in edge-coloured multigraphs, prove new ones and
consider several open problems on the topic. In Section
\ref{existsec} we briefly consider  a problem of whether an
edge-coloured graph  has a PC cycle. In Sections \ref{gadsec} and
\ref{gadgsec}, we offer a useful tool to study edge-coloured multigraphs.
In investigating problems on PC subgraphs of
edge-coloured multigraphs, it is convenient to transform an edge-coloured
graph into an ordinary graph. We suggest a new technique that
somewhat automates this transformation. Moreover, by proving some new
results, we illustrate how the proposed technique allows us to
obtain more efficient algorithms for PC  cycle and PC $(s,t)$-path
problems by reducing the order and size of the transformed graph. We
raise a problem of determining the minimum order and size of the
transformed graph, and describe the family of graphs that may be the solution to
the problem.

In Section \ref{longpcsec} we study long PC cycles and paths in
arbitrary edge-coloured multigraphs and Section \ref{completesec} is
devoted to longest (mostly Hamilton) PC cycles in edge-coloured
complete graphs.

An {\bf $m$-path-cycle subgraph} $F$ of a multigraph  $G$ is a
vertex-disjoint union of $m$ paths and a number of cycles in $G$
(some cycles can be of length 2). If
$m=0$, we call $F$ a {\bf cycle subgraph} of $G$. For a vertex set
$X$ of a multigraph  $G$, $G\la X\ra$ denotes the subgraph of $G$ induced
by $X.$ For a pair $s,t$ of distinct vertices of $G$, a path between
$s$ and $t$ is called an $(s,t)$-{\bf path}.

We consider {\bf edge-coloured multigraphs}, i.e., undirected multigraphs in
which each edge has a colour, but no parallel edges have the same colour.
If an edge-coloured multigraph  $G$ has $c$
colours, we assume that the colours are $1,2,\ldots, c$ and we call
$G$ a $c$-{\bf edge-coloured} multigraph. We denote the colour of an edge
$e$ of an edge-coloured multigraph  $G$ by $\chi(e).$ When $G$ has no parallel edges, we
call $G$ an {\bf edge-coloured graph}.

Let $G$ be a $c$-edge-coloured multigraph and let $v\in V(G)$. By
$N_i(v)$ we denote the set of neighbours of $v$ adjacent to $v$ by
an edge of colour $i$; let $d_i(x)=|N_i(x)|$.  The {\bf maximum}
({\bf minimum}) {\bf monochromatic degree} of $G=(V,E)$ is defined
by \begin{eqnarray*}\Delta_{mon}(G)&=&\max \{d_j(v):\ v\in V,\ 1\le
j\le c\}\\ (\delta_{mon}(G)&=&\min \{d_j(v):\ v\in V,\ 1\le j\le
c\}).\end{eqnarray*}  Let $\chi(v)=\{i:\ 1\le i\le c, N_i(v)\neq
\emptyset\}$. A path or cycle $Q$ of $G$ is {\bf properly coloured
(PC)} if every two adjacent edges of $Q$ are of different colours.

\section{Existence of PC Cycles}\label{existsec}

Since a pair of parallel edges in a
$c$-edge-coloured multigraph $(c\ge 2$) forms a PC cycle,
in this section, we consider only $c$-edge-coloured graphs.

It is easy to see that the problem of checking whether a
$c$-edge-coloured graph  has a PC cycle is more general (even for
$c=2$) than the simple problem of verifying whether a digraph
contains a directed cycle. Indeed, consider a digraph $D$ and, to
obtain a 2-edge-coloured graph  $G$ from $D$, replace each arc $xy$
of $D$ with edges $xz_{xy}$ and $z_{xy}y$ of colours 1 and 2, where
$z_{xy}$ is a new vertex ($z_{xy}\neq z_{x'y'}$ provided $xy\neq
x'y'$). Observe that $G$ has a PC cycle if and only if $D$ has a
directed cycle.

The following theorem by Yeo \cite{yeoJCTB69} provides a simple
recursive way of checking whether a $c$-edge-coloured graph  has a PC
cycle. (For $c=2$, Theorem \ref{GHtheorem} was first proved by
Grossman and  H\"aggkvist \cite{grossmanJCT34}.)

\begin{theorem}
\label{GHtheorem} Let $G$ be a $c$-edge-coloured graph, $c\ge 2$,
with no PC cycle. Then, $G$ has a vertex $z\in V(G)$ such that no
connected component of $G-z$ is joined to $z$ with edges of more
than one colour.
\end{theorem}

Let us consider the following function introduced by Gutin
\cite{gutinAJC}: $d(n,c)$, the minimum number $k$ such that every
$c$-edge-coloured graph  of order $n$ and minimum monochromatic
degree at least $k$ has a PC cycle. It was proved in
\cite{gutinAJC} that $d(n,c)$ exists and that
\begin{equation}\label{dncupperb} d(n,c)\le {1 \over {\lfloor c/2 \rfloor}} (\log_2 n
-{1 \over 3} \log_2\log_2 n + \Theta(1)).\end{equation}
Abouelaoualim et al. \cite{abou2007b} stated a conjecture which
implies that $d(n,c)=1$ for each $c\ge 2$. Using a recursive
construction inspired by Theorem \ref{GHtheorem} of
$c$-edge-coloured graphs with minimum monochromatic degree $p$ and
without PC cycles, Gutin \cite{gutinAJC} showed that
\begin{equation}\label{din} d(n,c)\ge {1 \over c}(\log_cn
-\log_c\log_cn)\end{equation} and, thus, the conjecture does not
hold. The bounds (\ref{dncupperb}) and  (\ref{din}) imply that
$d(n,c)=\Theta(\log_2 n)$ for every fixed $c\ge 2$.

\begin{conjecture}\cite{gutinAJC}
There is a function $s(c)$ dependent only on $c$ such that
$d(n,c)=s(c)\log_2 n(1+o(1))$.
\end{conjecture}

In particular, it would be interesting to determine $s(2)$.

\section{P-Gadgets}\label{gadsec}

We consider gadget constructions which generalize some known
constructions mentioned below. The P-gadget graphs $G^*$ and
$G^{**}$ of an edge-coloured multigraph  $G$ described in the next section
allow one to transform several problems on properly coloured
subgraphs of $G$ into perfect matching problems in $G^*$ or
$G^{**}$.

Let $G$ be an edge-coloured multigraph  and let $G'=G-\{x\in V(G):\
|\chi(x)|=1\}$. For each $x\in V(G')$ let $G_x$ be an arbitrary
(non-edge-coloured) graph  with the following four properties:

\begin{description}
  \item[P1] $\{x_q:\ q\in \chi(x)\}\subseteq V(G_x)$;
  \item[P2] $G_x$ has a perfect matching;
  \item[P3] For each $p\neq q\in \chi(x)$, if the graph
  $G_x-\{x_p,x_q\}$ is not empty, it has a perfect matching;
  \item[P4] For each set $L\subseteq \chi(x)$ with at least 3
  elements; if the graph
  $G_x-\{x_l:\ l\in L\}$ is not empty, it has no perfect matching.
  \end{description}

Each $G_x$ with the properties P1-P4 is called a {\bf P-gadget}. Let
us consider the following three P-gadgets; the first two are known
in the literature and the third one is new.

\begin{enumerate}

\item One P-gadget is due to Szeider \cite{szeiderDAM126}: $$V(G_x)=\{x_i,x'_i:\ i\in \chi(x)\}\cup
\{x''_a,x''_b\} \mbox{ and }$$
$$E(G_x)=\{x''_ax''_b,x_i'x_a'',x_i'x_b'':\ i\in \chi(x)\}\cup \{x_ix'_i:\ i\in
\chi(x)\}.$$ We will call this the {\bf SP-gadget}.

\item Another gadget is due to Bang-Jensen and Gutin \cite{bangDM165}:
$$V(G_x)=\{x_j:\ j\in \chi(x)\}\cup \{y_j:\ j\in \chi(x)\setminus
\{m,M\}\},$$ where $m=\min \chi(x),\ M=\max \chi(x)$, and
$$E(G_x)=\{x_jy_k:\ j\in \chi(x),\ k\in \chi(x)\setminus\{m,M\}\}\cup
\{x_jx_k:\ j\neq k\in \chi(x)\}.$$ We will call this the {\bf
BJGP-gadget}.

\item The following new gadget is a sort of crossover of the above two
and is called the {\bf XP-gadget}: $$V(G_x)=\{x_j:\ j\in
\chi(x)\}\cup \{y_j:\ j\in \chi(x)\setminus \{m,M\}\},$$ where $m$
and $M$ are defined above, and $$E(G_x)=\{x_mx_M\}\cup \{x_jy_j,\
x_my_j,\ x_My_j:\ j\in \chi(x)\setminus \{m,M\}\}.$$
\end{enumerate}

It is not difficult to verify that the tree P-gadgets indeed satisfy
P1-P4. Let $z=\chi(x)$. Observe that the SP-gadget has $2z+2$
vertices and $3z+1$ edges, the BJGP-gadget $2z-2$ vertices and
$z(3z-5)/2$ edges, the XP-gadget $2z-2$ vertices and $3z-5$ edges.
Thus, the XP-gadget has the minimum number of vertices and edges
among the three P-gadgets. It is not difficult to verify that the
XP-gadget has the minimum number of vertices and edges among  all
possible P-gadgets for $z=2,3,4$. Perhaps, this is true for any $z.$

\begin{conjecture}
The XP-gadget has the minimum number of vertices and edges among all
possible P-gadgets for every $z\ge 2$.
\end{conjecture}

We will see in the next section why minimizing the numbers of
vertices and edges in P-gadgets is important for speeding up some
algorithms on edge-coloured multigraphs.

\section{P-gadget Graphs}\label{gadgsec}

Let $G$ be a $c$-edge-coloured multigraph  and let $G_x$ be a P-gadget for
$x\in V(G')$. The graph  $G^*$ is defined as follows:
$V(G^*)=\cup_{x\in V(G')}V(G_x)$ and $E(G^*)=E_1\cup E_2$, where
$E_1=\cup_{x\in V(G')}E(G_x)$ and $E_2= \{y_qz_q:\ y,z\in V(G'),\
yz\in E(G),\ \chi(yz)=q,\ 1\le q\le c\}.$

Let $s,t$ be a pair of distinct vertices of $G$ and let
$H=G-\{s,t\}$. Let $G^{**}$ be constructed from $H^*$ by adding $s$
and $t$ and edges $E_3=\{sx_i:\ sx\in E(G), \chi(sx)=i\}\cup
\{tx_i:\ tx\in E(G), \chi(tx)=i\}.$

We will denote the number of vertices and edges in multigraph s $G$, $G^*$
and $G^{**}$ by $n,m,n^*,m^*,n^{**}$ and $m^{**}$, respectively.

The following result relates perfect matchings of $G^*$ with PC
cycle subgraphs of $G$.  PC cycle subgraphs are important in several
problems on edge-coloured multigraphs (for example, for the PC Hamilton
cycle problem), see \cite{bang2000}. Recall that $G'=G-\{x\in V(G):\
|\chi(x)|=1\}$.

\begin{theorem}\label{facttheorem}
Let $G$ be a connected edge-coloured multigraph  such that $G'$ is
non-empty. Then $G$ has a PC cycle subgraph with $r$ edges if and
only if $G^*$ has a perfect matching with exactly $r$ edges in
$E_2$.
\end{theorem}
\pf Let $M$ be a perfect matching of $G^*$ with exactly edges
$$x^1_{p_1}y^1_{q_1},\ldots , x^r_{p_r}y^r_{q_r}$$ in $E_2$. For
a vertex $x$ of $G'$, let $Q_x$ be the set of edges in $E_2$
adjacent to $G_x.$ By P2, each $G_x$ has even number of vertices
($x\in V(G')$) and since $M$ is a perfect matching in $G^*$, there
is even number of edges in $Q_x$. By P4, $Q_x$ has either no edges
or two edges for each $x\in V(G')$. Let $X$ be the set of all
vertices $x\in V(G')$ such that $|Q_x|=2$. Then, by the definition
of $G^*$, $G\la X\ra$ contains a PC cycle factor. It remains to
observe that $|X|=r.$

Now let $F$ be a PC cycle subgraph of $G$ with $r$ edges. Observe
that the edges of $F$ correspond to a set $Q$ of $r$ independent
edges of $G^*$ and that either no edges or two edges of $Q$ are
adjacent to $G_x$ for each $x\in V(G')$. Now delete the vertices
adjacent with $Q$ from each $G_x$ and observe that each remaining
non-empty gadget has a perfect matching by P2 and P3. Combining the
perfect matchings of the non-empty gadgets with $Q$, we get a
perfect matching of $G^*$ with exactly $r$ edges from $E_2$.\qed

\2

The first part of the next assertion generalizes a result from
\cite{bangDM165}. The second part is based on an approach which
leads to a more efficient algorithm than in \cite{abouLATIN2008}.

\begin{corollary}\label{maxpccs}
One can check whether an edge-coloured multigraph  $G$ has a PC cycle and,
if it does, find a maximum PC cycle subgraph of $G$ in time
$O(n^*\cdot (m^*+n^*\log n^*))$. Moreover one can find a shortest PC
cycle in $G$ in time $O(n\cdot n^*\cdot (m^*+n^*\log n^*))$.
\end{corollary}
\pf We may assume that $G$ is connected and that $G'$ is not empty.
By Theorem \ref{facttheorem}, it is enough to find a perfect
matching of $G^*$ containing the maximum number of edges from $E_2$.
Assign weight 0 (1, respectively) to edges of $G^*$ in $E_1$ ($E_2$,
respectively). Now we need to find a maximum weight perfect matching
of $G^*$ which can be done in time $O(n^*\cdot (m^*+n^*\log n^*))$
by a matching algorithm in \cite{gabowSODA1990}.

To find a shortest PC cycle in $G$, choose a vertex $x\in V(G')$. We
will find a shortest PC cycle in $G$ traversing $x$. By Theorem
\ref{facttheorem}, it is enough to find a perfect matching of $G^*$
containing the minimum number of edges from $E_2$ while containing
at least one edge from $E_2$ so that the corresponding PC cycle in
$G$ should be non-trivial. We define the weights on edges of $G^*$
as follows. Assign $M$, where $M$ is a sufficiently large number, to
each edge in $E_2$ incident with $G_x$. For all other edges, assign
weight 1 (0, respectively) to edges of $G^*$ in $E_1$ ($E_2$,
respectively). A maximum weight perfect matching of $G^*$ contains
exactly two edges of weight $M$ by P4, and contains the minimum
number of edges in $E_2$. Finding a maximum weight perfect matching
of $G^*$ can be done in time $O(n^*\cdot (m^*+n^*\log n^*))$ and we
iterate the process for each $x\in V(G')$.\qed

\2

The proof of the following result is analogous to the proof of
Theorem \ref{facttheorem}.

\begin{theorem}
Let $G$ be an edge-coloured multigraph  and let $s,t$ be a pair of
distinct vertices of $G$. If $G^{**}$ is non-empty, then $G$ has a
PC 1-path-cycle subgraph with  $r$ edges in which the path is
between $s$ and $t$ if and only if $G^{**}$ has a perfect matching
with exactly $r$ edges not in $E_1.$
\end{theorem}

The next assertion generalizes a result from \cite{abouLATIN2008}.

\begin{corollary}
Let $G$ be an edge-coloured multigraph.
One can check whether there is a PC $(s,t)$-path in $G$ in time
$O(m^{**})$ and if $G$ has one, a shortest PC $(s,t)$-path can be
found in time $O(n^{**}\cdot (m^{**}+n^{**}\log n^{**}))$.
\end{corollary}
\pf Let $L$ be a graph. Given a matching $M$ in $L$, a path $P$ in
$L$ is {\bf $M-$augmenting} if, for any pair of adjacent edges in
$P$, exactly one of them belongs to $M$ and the first and last edges
of $P$ do not belong to $M$. Consider a perfect matching $M$ of
$H^*$, where $H=G-\{s,t\}$, which is a collection of perfect
matchings of $G_x$ for all $x\in V(G')$. The existence of a perfect
matching in $G_x$ is guaranteed by P2. Observe that $G$ has a PC
$(s,t)$-path if and only if there is an $M-$augmenting $(s,t)$-path
$P$ in $G^{**}$. Since an $M-$augmenting path $P$ can be found in
time $O(m^{**})$ (see \cite{tarjan1983}), we can find a PC
$(s,t)$-path in $G$, if one exists, in time $O(m^{**})$.

To find a shortest PC $(s,t)$-path, we assign each edge in
$\bigcup_{x\in V(G')}E(G_x)$ weight 0 and every other edge of
$G^{**}$ weight 1. Observe that a minimum weight perfect matching
$Q$ in the weighted $G^{**}$ corresponds to a shortest PC
$(s,t)$-path. Finding a minimum weight perfect matching can be done
in time $O(n^{**}\cdot (m^{**}+n^{**}\log n^{**}))$. \qed

\section{Long PC Cycles and Paths}\label{longpcsec}

The following interesting result and conjecture were obtained by
Abouelaoualim, Das, Fernandez de la Vega, Karpinski, Manoussakis,
Martinhon and Saad \cite{abou2007b}.

\begin{theorem}\label{norn-1th} \cite{abou2007b}
Let $G$ be a $c$-edge-coloured multigraph $G$ with $n$ vertices and
with $\delta_{mon}(G)\ge \lceil \frac{n+1}{2} \rceil.$ If $c\ge 3$
or $c=2$ and $n$ is even, then $G$ has a Hamilton PC cycle. If $c=2$
and $n$ is odd, then $G$ has a PC cycle of length $n-1.$
\end{theorem}

\begin{conjecture} \cite{abou2007b}
Theorem \ref{norn-1th} holds if we replace $\delta_{mon}(G)\ge
\lceil \frac{n+1}{2} \rceil$ by  $\delta_{mon}(G)\ge \lceil
\frac{n}{2} \rceil$.
\end{conjecture}

We cannot replace $\delta_{mon}(G)\ge \lceil \frac{n+1}{2} \rceil$
by $\delta_{mon}(G)\ge \lceil \frac{n-1}{2} \rceil$ due to the
following example. Let $H_1$ and $H_2$ be $c$-edge-coloured complete
multigraphs (for each pair $x,y$ of vertices and each $i\in
\{1,2,\ldots ,c\}$ and $j\in \{1,2\}$, $H_j$ has a edge between $x$
and $y$ of colour $i$) of order $p+1$ that have precisely one vertex
in common. Clearly, a longest PC cycle in $H_1\cup H_2$ is of length
$p+1$.

Since the longest PC path problem is ${\cal NP}$-hard, it makes
sense to study lower bounds on the length of a longest PC path. The
following result was proved by Abouelaoualim et al.
\cite{abou2007b}.

\begin{theorem}\label{PCpathADFKMMSth}
Let $G$ be a $c$-edge-coloured graph of order $n$ with
$\delta_{mon}(G)=d\ge 1$. Then $G$ has a PC path of length at least
$\min\{n-1,2\lfloor \frac{c}{2}\rfloor d\}$.
\end{theorem}

The authors of \cite{abou2007b} raised the following two
conjectures.

\begin{conjecture}
Let $G$ be a $c$-edge-coloured graph  of order $n$ and let
$d=\delta_{mon}(G)\ge 1$. Then $G$ has a PC path of length at least
$\min\{n-1,2cd\}$.
\end{conjecture}

They also conjectured the following analog of Theorem
\ref{PCpathADFKMMSth} for multigraphs:

\begin{conjecture}
Let $G$ be a $c$-edge-coloured multigraph of order $n$ with
$\delta_{mon}(G)=d\ge 1$. Then $G$ has a PC path of length at least
$\min\{n-1,2d\}$.
\end{conjecture}

\section{Longest PC Cycles and Paths in Edge-Coloured Complete
Graphs}\label{completesec}

Let $K^c_n$  denote a $c$-edge-coloured complete graph  with $n$
vertices.

Feng, Giesen, Guo, Gutin, Jensen and Rafiey \cite{fengJGT53} proved the following:

\begin{theorem}\label{PCHPth}
A $K^c_n$ ($c\ge 2$) has a PC Hamilton path if and only if $K^c_n$
contains a PC spanning 1-path-cycle subgraph.
\end{theorem}

This theorem was first proved by Bang-Jensen and Gutin
\cite{bangDM165} for the case $c=2$ and they conjectured that Theorem \ref{PCHPth}
holds for each $c\ge 2$. Theorem \ref{PCHPth} implies
that the maximum order of a PC path in $K^c_n$ equals the maximum
order of a PC 1-path-cycle subgraph of $K^c_n$.

As a result, the problem of finding a longest PC path in $K^c_n$ is
polynomial-time solvable for arbitrary $c\ge 2$. To see that a PC
1-path-cycle subgraph of $K^c_n$ can be found in polynomial time,
add a pair $x,y$ of new vertices to $K^c_n$ together with all edges
needed to have a complete multigraph  on $n+2$ vertices. Let the colour of
all edges between $x$ and $y$ and $K_n^c$ be $c+1$ and let the
colour of $xy$ be $c+2.$ Observe that the maximum order of a PC
1-path-cycle subgraph of $K^c_n$ equals the maximum order of a PC
cycle subgraph of the $c+2$-edge-coloured complete graph  described
above. It remains to apply Corollary \ref{maxpccs}.

The problem of finding a longest PC cycle $K^c_n$ has not been
solved yet for $c\ge 3$ as we will see below. For $c=2$, Saad
\cite{saadCPC5} found a characterization for longest PC cycles using
the following notions. A pair of distinct vertices $x,y$ of $G$ are {\bf
colour-connected} if there exist PC $(x,y)$-paths $P$ and $Q$ such
that $\chi(f_P)\neq \chi(f_Q)$  and $\chi(\ell_P)\neq \chi(\ell_Q)$,
where $f_P$ and $f_Q$ are the first edges of $P$ and $Q$,
respectively, and $\ell_P$ and $\ell_Q$ are the last edges of $P$
and $Q$, respectively. We say that $G$ is {\bf colour-connected} if every pair
of distinct vertices of $G$ is colour-connected. Saad's characterization is
as follows.

\begin{theorem}
\label{Saadtheorem} The length of a longest PC cycle in a
colour-connected $K^2_n$ is equal to the maximum order of a PC cycle
subgraph of $K^2_n$.
\end{theorem}

Colour-connectivity for $K^c_n$ is an an equivalence relation (see
\cite{bang2000}). Using Theorem \ref{Saadtheorem}, Saad
\cite{saadCPC5} showed that the problem of finding a longest PC
cycle in $K^2_n$ is random polynomial. Using a special case of
Corollary \ref{maxpccs}, Bang-Jensen and Gutin \cite{bangDM188}
proved that the problem is, in fact, polynomial-time solvable.
Theorem \ref{Saadtheorem} implies the following:

\begin{corollary}\cite{saadCPC5}
\label{Saadcor} A $K^2_n$ has a PC Hamilton cycle if and only if
$K^2_n$ is colour-connected and contains a PC cycle factor.
\end{corollary}

There is another characterization of $K^2_n$ with a PC Hamilton
cycle due to Bankfalvi and Bankfalvi, see \cite{bang2000}. The
straightforward extension of Corollary \ref{Saadcor} is not true for
any $c\ge 3$ \cite{bang2000}. In fact, no characterization of
$K^c_n$ with a PC Hamilton cycle is known for any fixed $c\ge 3$ and
it is a very interesting problem to obtain such a characterization.
Even the following problem by Benkouar, Manoussakis, Paschos and
Saad \cite{benkouarLNCS557} is still open.

\begin{problem}
\label{HCcgeq3} Determine the complexity of the PC Hamilton cycle
problem for $c$-edge-coloured complete graphs when $c\geq 3$.
\end{problem}

We conjecture that the PC Hamilton cycle problem for
$c$-edge-coloured complete graphs when $c\geq 3$ is polynomial-time
solvable.

In absence of characterization of $K^c_n$ with a PC Hamilton cycle,
sufficient conditions are interest. Manoussakis, Spyratos, Tuza and
Voigt \cite{manoussakisGC12} proved the next result.

\begin{proposition}
\label{mincimpham}  If $c\ge \frac{1}{2}(n-1)(n-2)+2$, then every
$K^c_n$ has a PC Hamilton cycle.
\end{proposition}

Let  $\Delta_{mon}(K^c_n)$ denote the largest monochromatic degree
of $K^c_n$. Bollob\'as and Erd\H os \cite{bollobasIJM23} posed the
following:

\begin{conjecture}\label{BEconj}
 Every  $K^c_n$ with $\Delta_{mon}(K^c_n)\le
\lfloor n/2 \rfloor -1$ has a PC Hamilton cycle.
\end{conjecture}

Improving some previous results on this conjecture, Shearer
\cite{shearerDM25} showed that if $7\Delta_{mon}(K^c_n) < n$, then
$K^c_n$ has a PC Hamilton cycle. So far, the best asymptotic
estimate  was obtained by Alon and Gutin \cite{alonRSA11}.

\begin{theorem}\cite{alonRSA11}
\label{AGtheorem} For every $\epsilon>0$ there exists an
$n_0=n_0(\epsilon)$ so that for each $n>n_0$, every $K_n^c$
satisfying
$\Delta_{mon}(K_n^c) \leq (1-\frac{1}{\sqrt 2}
-\epsilon)n$
contains a PC Hamilton cycle.
\end{theorem}

\end{document}